\documentclass[conference,9pt]{IEEEtran}
\IEEEoverridecommandlockouts
\usepackage{spconf}
\usepackage{url}
\usepackage{subcaption} 
\usepackage{amsmath}
\usepackage{amssymb}
\usepackage{xcolor}
\usepackage{multirow}
\usepackage{etoolbox,balance}
\usepackage{cite}

\usepackage{balance}
\usepackage{pifont}
\usepackage{stmaryrd}
\usepackage{optidef}
\usepackage{mathrsfs}
\usepackage{amsfonts}  
\usepackage[bookmarks=false]{hyperref}

\usepackage{graphicx}
\usepackage{algorithm}
\usepackage{cuted}

\usepackage{algorithm}
\usepackage{algpseudocode}

\usepackage{booktabs}
\usepackage{dsfont}
\usepackage{esint}

\patchcmd{\thebibliography}{\section*{\refname}}{}{}{}

\title{Advancing Single-Snapshot DOA Estimation with Siamese Neural Networks \\ for Sparse Linear Arrays}

\name{Ruxin Zheng$^{\dagger}$, Shunqiao Sun$^{\dagger}$, Hongshan Liu$^{\dagger}$, and Yimin D.\ Zhang$^{\ddagger}$
\thanks{\hspace{-1.0em}The work of R.\ Zheng, S.\ Sun, and H.\ Liu was supported in part by National Science Foundation (NSF) under Grants CCF-2153386 and ECCS-2340029. The work of Y.\ D.\ Zhang was supported in part by NSF under Grant ECCS-2236023.}}
\address{$^{\dagger}$Department of Electrical and Computer Engineering, The University of Alabama, Tuscaloosa, AL, USA \\ 
$^{\ddagger}$Department of Electrical and Computer Engineering, Temple University, Philadelphia, PA, USA}

\begin{document}

\maketitle

\begin{abstract}
Single-snapshot signal processing in sparse linear arrays has become increasingly vital, particularly in dynamic environments like automotive radar systems, where only limited snapshots are available. These arrays are often utilized either to cut manufacturing costs or result from unintended antenna failures, leading to challenges such as high sidelobe levels and compromised accuracy in direction-of-arrival (DOA) estimation. Despite deep learning's success in tasks such as DOA estimation, the need for extensive training data to increase target numbers or improve angular resolution poses significant challenges. In response, this paper presents a novel Siamese neural network (SNN) featuring a sparse augmentation layer, which enhances signal feature embedding and DOA estimation accuracy in sparse arrays. We demonstrate the enhanced DOA estimation performance of our approach through detailed feature analysis and performance evaluation. The code for this study is available at \url{https://github.com/ruxinzh/SNNS_SLA}.

\end{abstract}

\begin{keywords}
Siamese neural network, single snapshot, direction-of-arrival estimation, sparse linear arrays
\end{keywords} 
\vspace{0.5em}
\section{Introduction}
Radar technology is crucial for the advancement of autonomous driving systems, offering robust performance even in adverse weather conditions \cite{sun2020mimo, markel2022radar,Ruxin_TAES_2023}. However, current automotive radar systems often struggle with low angular resolution due to the limited size of array apertures. Addressing the challenge of achieving substantial antenna aperture sizes for improved angular resolution, especially in filled arrays that require numerous antennas, sparse linear arrays (SLAs) have emerged as an efficient and economical solution in automotive radars \cite{sun2020sparse, sun20214d, xu2023automotive,zheng2023time}. These arrays enable larger apertures and better angular resolution with fewer antenna elements and help reduce mutual coupling thanks to their wider inter-element spacing. However, the task of designing optimal sparse arrays is formidable, as the ideal configuration heavily depends on diverse, specific requirements, underscoring the absence of a one-size-fits-all solution for sparse array design \cite{zheng20234d,lin2022design}. Additionally, random sensor failures can result in unpredictably sparse array geometries, further complicating the design process.

Recently, deep learning strategies for direction-of-arrival (DOA) estimation have gained prominence \cite{papageorgiou2021deep,fuchs2022machine,feintuch2023neural,gall2020spectrum,gall2020learning,9827881,eamaz2024automotive}, known for rapid inference, enhanced super-resolution, and robust performance in low signal-to-noise ratio (SNR) environments \cite{papageorgiou2021deep}. These networks generate a pseudo-angle spectrum on a fixed grid, treating DOA estimation as a multi-label classification task. However, as the grid becomes finer, indicating a greater number of labels and potential targets, the number of label combinations grows exponentially, which exacerbates training data requirements and leads to label imbalance issues, thus potentially degrading the performance of such data-driven approaches. Additionally, many models oversimplify by assuming uniform target intensity, which can compromise their effectiveness and generalizability in real-world scenarios. Furthermore, these networks must address random sensor failures \cite{lee2019robust,keizer2007element,liu2018robustness,vigneshwaran2007direction} to maintain the reliability of automotive radar systems. 

This paper introduces a novel Siamese neural network (SNN) \cite{bromley1993signature} designed for single-snapshot DOA estimation in SLAs. SNNs, utilizing twin identical networks, are ideal for applications requiring similarity learning between data elements. Our approach employs SNNs to discern similarities between signals with identical DOAs but varying in reflection coefficients, sparse array geometries, and noise levels. This technique significantly improves the network's capability to encode data features accurately. We demonstrate the effectiveness of the proposed method through feature analysis and DOA estimation performance evaluations, highlighting the adaptability of the proposed method across different array configurations.

\vspace{0.5em}
\section{System Model}\label{sigm}
Consider a linear antenna array with $N$ omnidirectional elements receiving $K$ narrowband, far-field signals from sources $s_k(t)$ arriving from directions $\theta_k$ for $k = 1,2, \cdots, K$. The received signals at different sensors manifest as phase shifts, leading to the data model:
\begin{equation}
\mathbf{y}(t) = \mathbf{A}(\boldsymbol  \theta) \mathbf{s}(t) + \mathbf{n}(t), \quad t = 1, \cdots, T,
\label{eq1}
\end{equation}
where $\mathbf{n}(t)$ is the $N \times 1$ complex white Gaussian noise vector and $\mathbf{A}(\boldsymbol  \theta) = [\mathbf{a}(\theta_1), \dots, \mathbf{a}(\theta_K)]$ is the $N \times K$ array manifold matrix. The steering vector $\mathbf{a}(\theta_k)$ for direction $\theta_k$ is defined as
\begin{equation}
\mathbf{a}(\theta) = \left[1, e^{j\frac{2\pi d_2}{\lambda}\sin{\theta}}, \dots , e^{j\frac{2\pi d_N}{\lambda}\sin{\theta}}\right]^T,
\end{equation}
where $(\cdot)^{T}$ denotes transpose, $d_n$ stands for the spacing between the first element and the $n$-th element for $n=2, 3, \cdots, N$, and $\mathbf{s}(t) = [s_1(t), s_2(t), \dots, s_K(t)]^T$ represents the source signals. Because this study analyzes the single-snapshot response $\mathbf{y}$ of the array, we omit the time variable $t$, and Eq.\ (\ref{eq1}) becomes
\begin{equation}
\mathbf{y} = \mathbf{A}(\boldsymbol  \theta)\mathbf{s} + \mathbf{n}.
\end{equation}

\begin{figure}
    \centering
    \begin{subfigure}[b]{0.4\textwidth} 
        \centering
        \includegraphics[width=\textwidth]{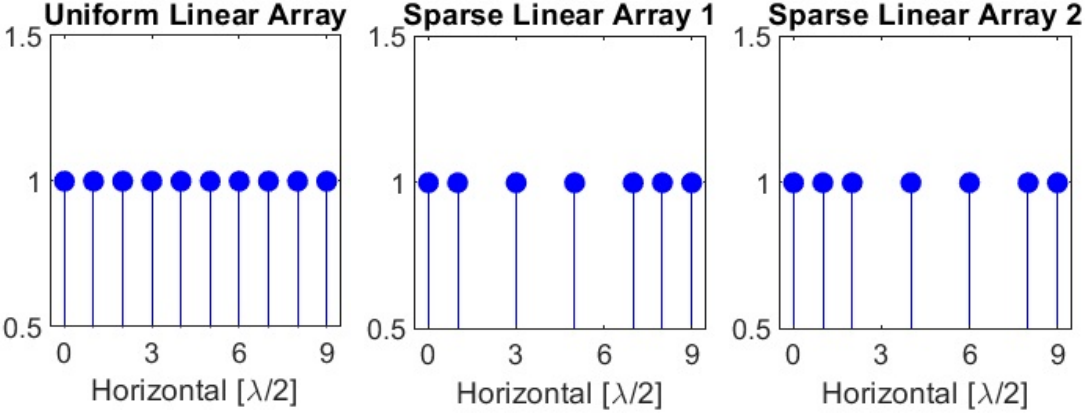} 
        \caption*{(a)} 
    \end{subfigure}
    \begin{subfigure}[b]{0.4\textwidth} 
        \centering
        \includegraphics[width=\textwidth]{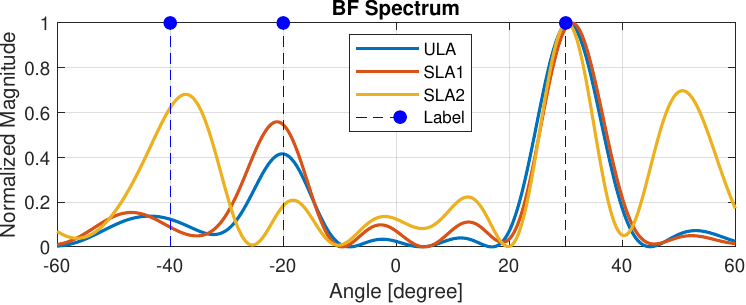}
        \caption*{(b)} 
    \end{subfigure}
    \caption{(a) Example of ULA and SLAs. The SLAs have a 0.3 sparsity. (b) The beamforming spectrum of a single-snapshot signal at 20 dB SNR, depicting three targets with different amplitudes and DOAs, consistently labeled across various array configurations.}\label{ULASLA}
    \label{fig:test}
\end{figure}

Figure \ref{ULASLA}(a) compares a 10-element uniform linear array (ULA) with inter-element spacing of half a wavelength and two 7-element SLAs, which maintain the same array aperture of $4.5\lambda$ as the ULA but use less antenna elements modeled by a binary mask, with $\lambda$ denoting the signal wavelength. The sparsity of the SLAs is defined as
\begin{equation}
\text{Sparsity} = 1 - \frac{N_{\text{SLA}}}{N_{\text{ULA}}},
\end{equation}
where $N_{\text{ULA}}$ and $N_{\text{SLA}}$ respectively represent the number of antennas in the ULA and the SLA. For both SLA examples shown in Figure \ref{ULASLA}(a), the sparsity is 0.3, indicating a $30\%$ reduction in the number of elements compared to the ULA.

Figure \ref{ULASLA}(b) displays the beamforming (BF) spectrum of a single-snapshot signal for the ULA and the two SLAs, 
when the beam is steered toward the $30^{\circ}$ direction. 
The results feature a 20 dB SNR and three targets with reflectivity intensities of 0.2, 0.5, and 1, and DOAs of $-40^{\circ}$, $-20^{\circ}$, and $30^{\circ}$, respectively. Despite variations in the spectra from different array configurations, all are assigned the same label for supervised learning purposes, reflecting their fundamental similarity. This consistency underpins our decision to use SNNs for processing signals in SLAs, capitalizing on their ability to learn and leverage these similarities effectively.

\vspace{0.5em}
\section{Network Design and Architecture}
\subsection{Challenges}

Treating DOA estimation as a multilabel classification task, the size of the network output, \(N_{\text{out}}\), depends on the granularity of the angle grid.  For a field of view between $-60^{\circ}$ and $60^{\circ}$ at a $1^{\circ}$ interval, the output size \(N_{\text{out}}\) is 121. The total number of possible labels, \(N_{\text{labels}}\), given the number of targets, \(N_{\text{targets}}\), is
\begin{equation}
N_{\text{labels}} = \sum_{k=1}^{N_{\text{targets}}} \binom{N_{\text{out}}}{k}.
\end{equation}
Even without considering variations in the input SNR, target reflectivity, and diverse sparse array geometries, increasing \(N_{\text{out}}\) and \(N_{\text{targets}}\) alone would exponentially inflate the training data requirement \cite{papageorgiou2021deep}, posing challenges like severe class imbalance and making the model nearly untrainable with a vast but underrepresented label spectrum.

\begin{figure} [t]
\centering
\includegraphics[width=0.45\textwidth]{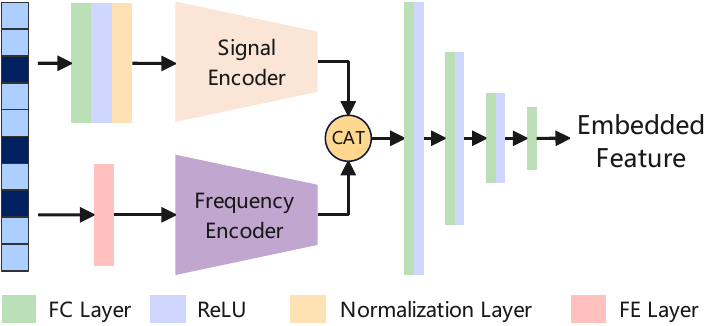}
\caption{Encoder architecture for feature extraction, where `CAT' indicates vector concatenation.}
\label{encoder}
\end{figure}

\subsection{Network Architecture}
To overcome these challenges, we propose a novel SNN featuring a specially designed sparse augmentation layer and frequency embedding layer, tailored for DOA estimation of single-snapshot signals in SLAs.
\begin{figure*} 
\centering
\includegraphics[width= 0.65\textwidth]{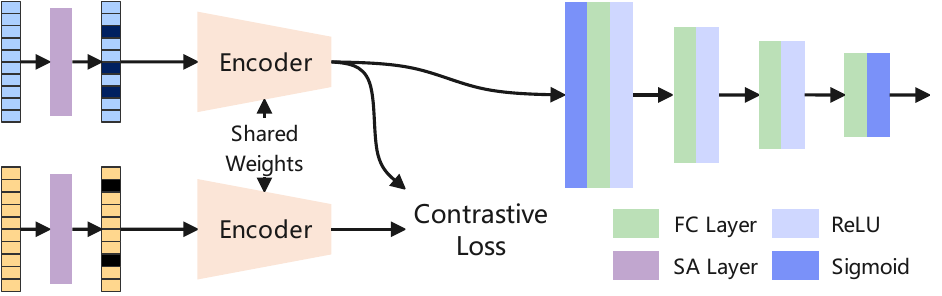}
\caption{Network architecture of the proposed SNN with sparse augmentation layer.}\label{SNN}
\end{figure*}

\textbf{\textit{Sparse Augmentation (SA) Layer:}}
Data augmentation is crucial for enhancing deep learning model robustness and preventing overfitting. This is implemented by artificially expanding the dataset with transformations such as flipping, rotation, and translation \cite{shorten2019survey}.

In signal processing, the sparse augmentation layer specifically introduces controlled sparsity into the dataset. This layer generates a random binary mask matching the input signal size and includes a configurable parameter for maximum sparsity as detailed in Section \ref{sigm}. For example, setting the maximum sparsity to 0.3 in a 10-element ULA allows zeroing out up to three elements, forming a sparse representation. The layer also outputs the count of activated antenna elements, which is used for normalization. During training, this count is determined by the sparse augmentation layer, while in evaluation, it is set through thresholding algorithms.

\textbf{\textit{Frequency Embedding (FE) Layer:}} Incorporating domain knowledge through handcrafted features significantly enhances the performance of deep learning networks, especially in complex or poorly understood domains. The sparse augmentation layer, which randomly masks signals, makes the input unsuitable for convolutional layers. To remedy this, we introduce the sparse signal frequency embedding layer that transforms sparse signals into a continuous frequency domain, facilitating the application of convolutional layers for feature extraction. This transformation is mathematically defined as:
\begin{equation}
g(\mathbf{y}) = \frac{\mathbf{A}^H(\boldsymbol \theta) \mathbf{y}}{N_{\text{SLA}}},
\end{equation}
where 
$(\cdot)^H$ denotes conjugate transpose and $\mathbf{y}$ stands for the input signal vector. 
This layer allows for effective embedding of sparse signals, preserving data integrity for enhanced feature extraction.

\textbf{\textit{Siamese Neural Network:}}\label{snn}
Figure \ref{encoder} illustrates the architecture of our feature encoder. Initially, a sparse signal is processed through a fully connected (FC) layer and a ReLU activation layer, mathematically expressed as
\begin{equation}
  f(\mathbf{x}) = \sigma (\mathbf{W}\mathbf{x} + \mathbf{b}),  
\end{equation}
where \( \mathbf{W} \) and \( \mathbf{b} \) represent the weight matrix and bias vector, respectively, and $\sigma(\cdot)$ denotes the ReLU activation operation. A normalization layer, \( h(\mathbf{x}) = {\mathbf{x}}/{N_{\text{SLA}}} \), where \( N_{\text{SLA}} \) counts the non-zero elements, stabilizes the output features. This ensures consistency despite varying sparsity by adjusting features relative to active inputs, thus enhancing model reliability. The normalized signal is then fed into a signal encoder, which consists of four FC layers, each followed by a ReLU layer, to extract signal features. Concurrently, the sparse signal is processed through a frequency embedding layer, followed by a frequency encoder. This encoder is structured as a convolutional neural network (CNN) with four convolutional layers, where each layer is accompanied by a ReLU and max pooling layer to extract frequency features. Finally, both signal and frequency features are concatenated and processed through four additional FC layers with ReLU activations to form the embedded feature.

Figure \ref{SNN} illustrates the architecture of the proposed SNN. During training, the SNN processes two inputs. Input signals are classified as similar pairs if they have identical target DOAs but differ in target reflection coefficients and SNRs. Conversely, dissimilar pairs have different target DOAs. Each signal first passes through the SA layer, which randomly masks values to create sparse signals. Notably, after the SA layer, signals defined as similar not only vary in reflection coefficients and SNRs but also exhibit distinct sparse array configurations. These signals are subsequently fed into an encoder that extracts features, which are evaluated using a contrastive loss function to refine the embedding space. The function aims to minimize the distance between similar pairs while ensuring that dissimilar pairs are separated by at least a specified margin. The contrastive loss is mathematically expressed as
\begin{equation}
S = \frac{1}{P} \sum_{j=1}^{P} \left( z_j g_j^2 + (1 - z_j) \max(0, m - g_j)^2 \right),
\end{equation}
where \( g_j = \| v_{j1} - v_{j2} \| \) denotes the Euclidean distance between the paired samples \( v_{j1} \) and \( v_{j2} \). Here, \( v_{j1} \) and \( v_{j2} \) represent the embedded representations of two signal features, and \( P \) is the total number of these paired samples included in the computation. The binary label \( z_j \) indicates whether the pair is similar (\( z_j = 1 \)) or dissimilar (\( z_j = 0 \)). The parameter \( m \) is the margin, which sets a threshold for the distance below which pairs are penalized if considered dissimilar. This margin helps to regulate the influence of the distance on the learning process, promoting an optimal separation between similar and dissimilar pairs in the feature space.

Additionally, a binary cross-entropy loss is utilized for the multilabel classification task. The total loss for the SNN is a linear combination of the contrastive loss and binary cross-entropy loss. 

\subsection{Data Generation and Labeling}
We simulate signals using a $20$-element ULA with a half-wavelength inter-element spacing. The simulations target a maximum of $3$ distinct sources, each separated by at least $\Delta \phi = 1^{\circ}$. 
The field of view is defined as $\boldsymbol{\phi}_{\mathrm{FOV}} = [-60^{\circ}, 60^{\circ}]$, discretized at $1^{\circ}$ intervals, forming a directional grid $G \in \mathbb{R}^{1 \times M}$ with $M = 121$ possible DOAs. A random complex reflection coefficient $s$ is assigned to each DOA, with its magnitude uniformly distributed between $0.5$ and $1$. Signals are labeled using the ground truth as
\begin{equation}\label{bvgt}
  {\bf GT}_{n} =
    \begin{cases}
      1, & \text{if } \theta_k = G_n,\\
      0, & \text{otherwise}
    \end{cases} 
\end{equation}
for $n = 1, 2, \cdots, M$.
We generate $295,361$ unique angular combinations. For each combination, we produce $50$ signals, each with a randomly assigned input SNR ranging between $0$ dB and $30$ dB.

\subsection{Training Approach}
We implemented the proposed network and benchmark models using PyTorch, utilizing the Adam optimizer with a learning rate of \(1 \times 10^{-4}\). Training is proceeded for $1,000$ epochs with a batch size of 1,024, and is accelerated using four Nvidia RTX A6000 GPUs to enhance efficiency.

\vspace{0.5em}
\section{Performance Evaluation}\label{perf}
To more succinctly demonstrate the effectiveness of the SA layer and SNN, we employ two ``base network" benchmarks, namely ``BaseNet1,'' which is the proposed network without the SA layer and contrastive loss, and ``BaseNet2,'' which lacks only the contrastive loss. Additionally, we employ the traditional Compressive Sensing via Orthogonal Matching Pursuit (CS-OMP) algorithm\cite{tropp2007signal} as a conventional DOA estimation benchmark.

\subsection{Feature Analysis}
\begin{figure}[ht]
    \centering
    \begin{subfigure}[b]{0.48\textwidth} 
        \centering
        \includegraphics[width=\textwidth]{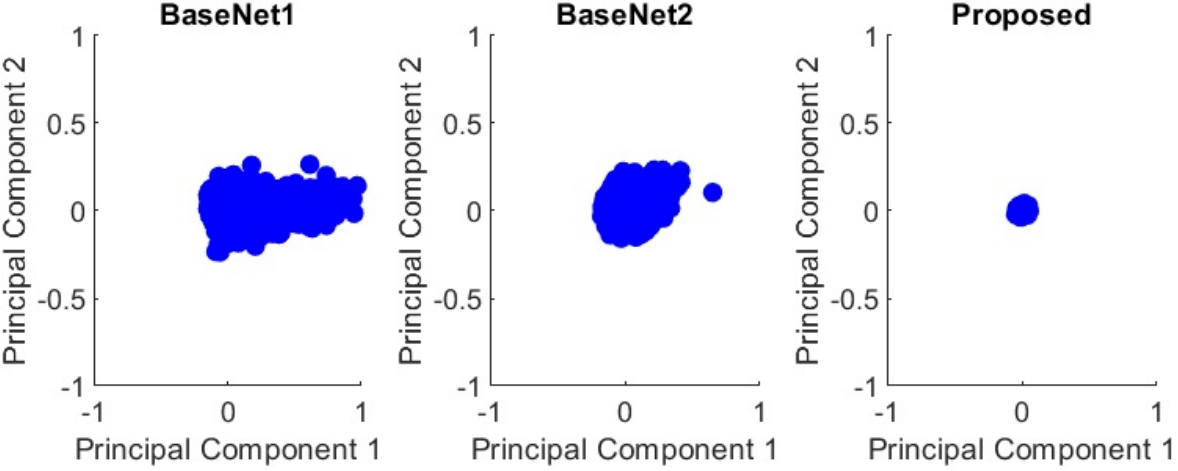} 
        \caption*{(a)} 
    \end{subfigure}
    \begin{subfigure}[b]{0.48\textwidth} 
        \centering
        \includegraphics[width=\textwidth]{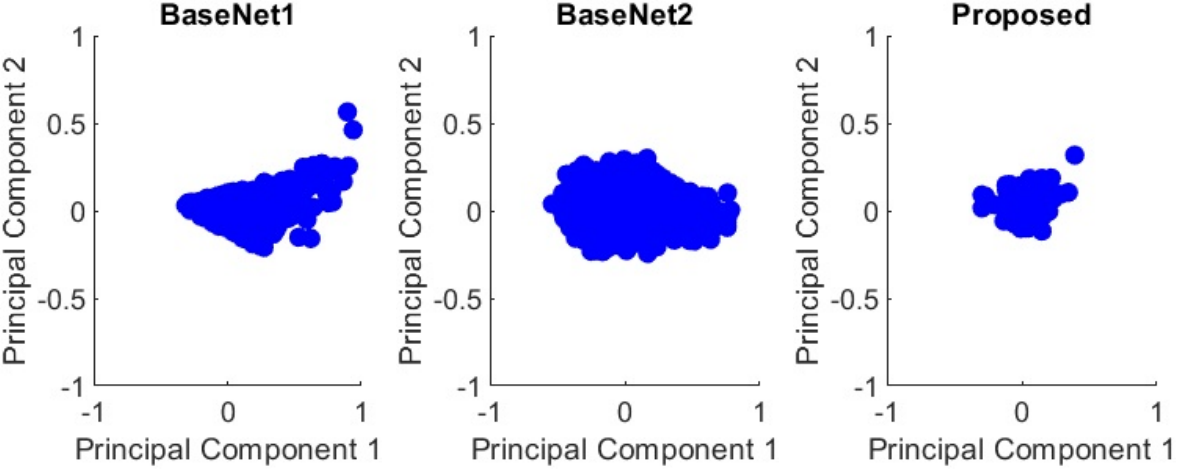}
        \caption*{(b)} 
    \end{subfigure}
    \caption{PCA-reduced embedding features for (a) ULA and (b) Random SLAs with a sparsity of 0.3.}\label{pca}
\end{figure}

\begin{figure}[ht]
    \centering
    \begin{subfigure}[b]{0.49\textwidth} 
        \centering
        \includegraphics[width=.49\textwidth]{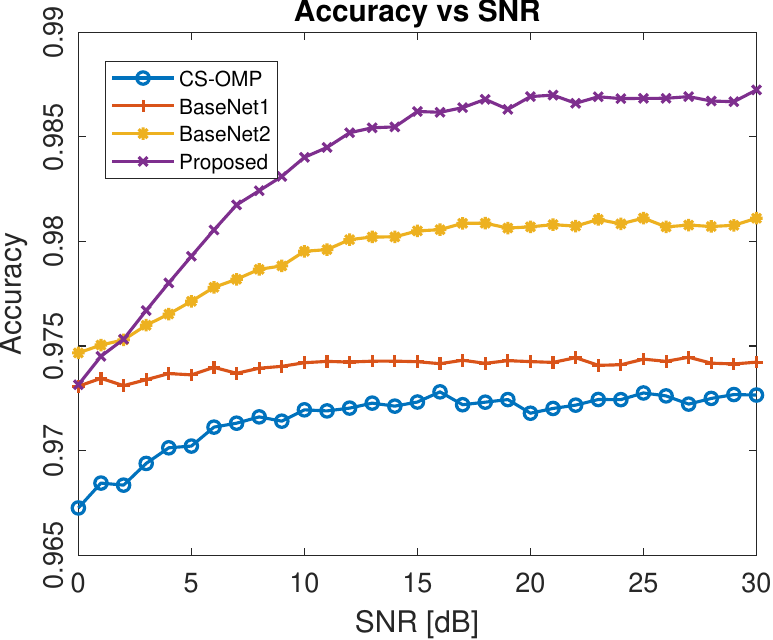} 
        \includegraphics[width=.49\textwidth]{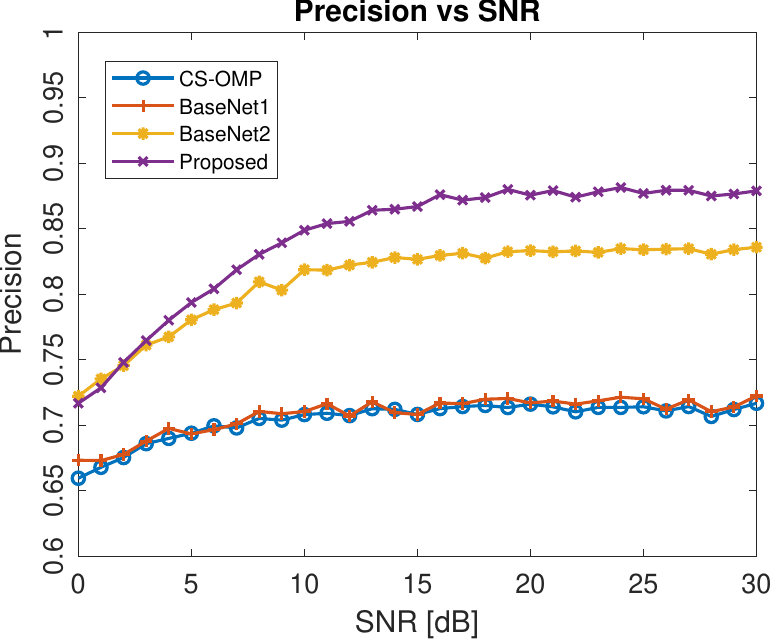} 
        \caption*{\quad (a) \hspace{13em} (b)} 
        \label{fig:sub1}
    \end{subfigure}
    \begin{subfigure}[b]{0.49\textwidth} 
        \centering
        \includegraphics[width=.49\textwidth]{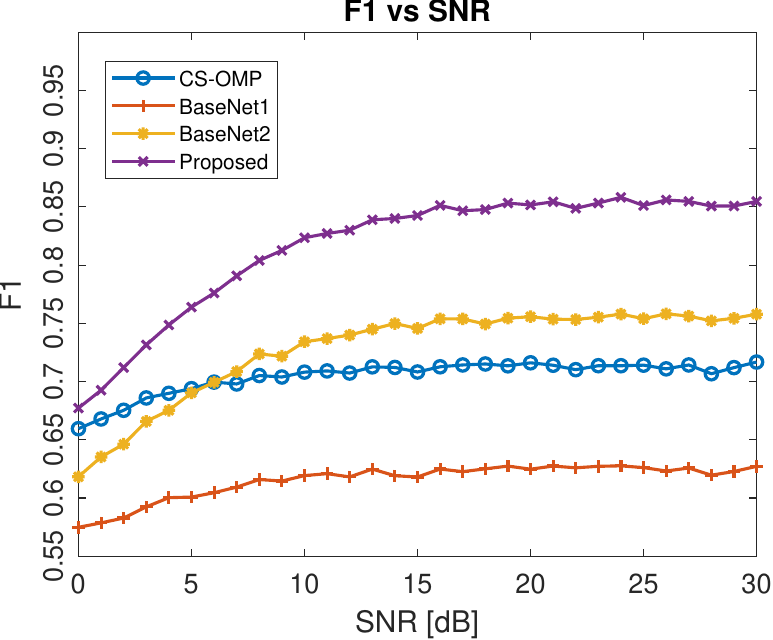} 
        \includegraphics[width=.49\textwidth]{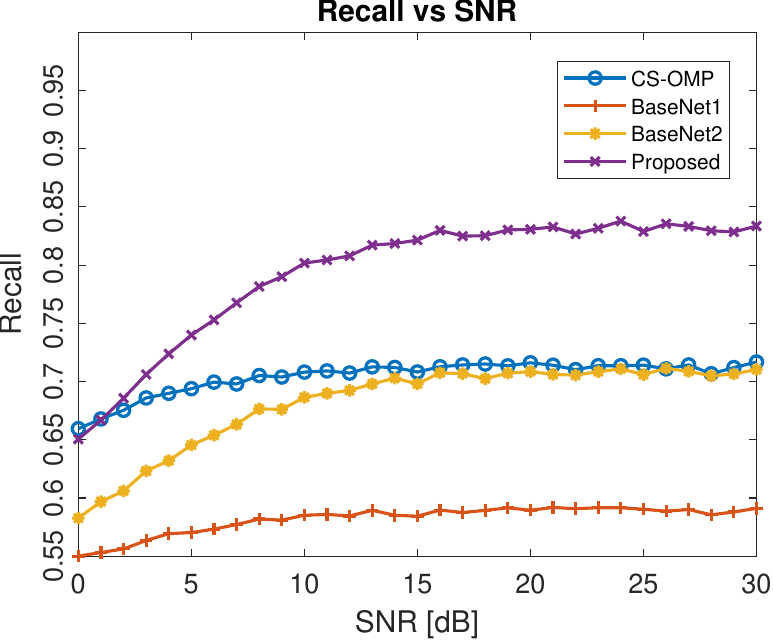} 
        \caption*{\quad (c) \hspace{13em} (d)} 
        \label{fig:sub3}
    \end{subfigure}
    \caption{Evaluation of DOA performance metrics: (a) Accuracy, (b) Precision, (c) F1-Score, and (d) Recall.}\label{doa}
\vspace{-1em}
\end{figure}

The feature representations extracted by the encoder from the same class should ideally be identical, or at least closely clustered. The proximity of these features directly correlates with improved classification performance. Therefore, in this subsection, we conduct a feature analysis. Since the output features from the encoder are high-dimensional, we utilize Principal Component Analysis (PCA) to reduce the dimensionality. This aids in better visualizing the relationships among the features. Figure \ref{pca} displays the PCA results for 5,000 signals that share the same DOA combinations but differ in noise levels and reflection coefficients. Figure \ref{pca}(a) reveals that BaseNet2, which has the SA layer, achieves notably tighter clustering of features compared to BaseNet1, which lacks this layer, highlighting the importance of feature embedding by the SA layer. The proposed SNN exhibits even tighter clustering, underscoring its advanced embedding capabilities. This indicates that the Siamese architecture, combined with the contrastive loss, significantly boosts the model’s ability to effectively discern and group similar features. Figure \ref{pca}(b) displays the embedded features of 5,000 test signals with randomly configured SLAs. To achieve a sparsity level of 0.3, six random positions in each test signal are set to zero. This figure illustrates the impact of signal sparsity on all three models. However, the proposed model demonstrates the least impact, benefiting from the robust integration of the Siamese architecture and the contrastive loss mechanism.

\subsection{DOA Estimation}

DOA estimation is treated as a multilabel classification task. For performance evaluation, we employ accuracy, precision, recall, and F1 score as the metrics. The evaluation involved $5,000$ test signals across random SLAs with a sparsity of 0.3, where the input SNR levels range between 0 dB and 30 dB, and the threshold is set to 0.5. 

The evaluation metrics are defined below, where \(TP_m\), \(TN_m\), \(FP_m\), and \(FN_m\) respectively represent the true positives, true negatives, false positives, and false negatives for the \(m\)th label for $m=1, \cdots, M$. 

\medskip
\noindent Accuracy: 
\begin{equation}
\text{Accuracy} = \frac{1}{M} \sum_{m=1}^M \frac{TP_m + TN_m}{TP_m + TN_m + FP_m + FN_m}.
\end{equation}

\medskip
\noindent Precision:
\begin{equation}
\text{Precision} = \frac{1}{M} \sum_{m=1}^M \frac{TP_m}{TP_m + FP_m}.
\end{equation}
\noindent Recall: 
\begin{equation}
\text{Recall} = \frac{1}{M} \sum_{m=1}^M \frac{TP_m}{TP_m + FN_m}.
\end{equation}

\medskip
\noindent F1-Score: 
\begin{equation}
\text{F1-Score} =  2 \cdot \frac{\text{Precision} \times \text{Recall}}{\text{Precision} + \text{Recall}}.
\end{equation}
\medskip

As illustrated in Figures \ref{doa}, BaseNet1 exhibits the poorest performance among the models evaluated, while BaseNet2 significantly surpasses BaseNet1 across all evaluation metrics at various SNR levels. This improvement is attributed to the SA layer, which equips BaseNet2 with the capability to effectively manage randomly sparsed input signals, thus demonstrating the effectiveness of the SA layer. Figures \ref{doa}(a) and \ref{doa}(b) further show that BaseNet2 and the proposed SNN outperform both CS-OMP and BaseNet1, underscoring the advantages of deep learning strategies over traditional methods.

Moreover, Figures \ref{doa}(c) and \ref{doa}(d) reveal that while BaseNet1 continues to lag in performance, the proposed method outstrips BaseNet2 owing to the integration of a Siamese architecture and contrastive loss, highlighting its enhanced robustness and efficacy. Overall, the proposed SNN achieves superior performance across all four evaluation metrics at different SNR levels, demonstrating the effectiveness and exceptional performance of our proposed approach.

\vspace{0.5em}
\section{Conclusion}
In conclusion, this study successfully addressed challenges in sparse linear arrays, particularly in automotive radar systems with limited snapshots. We developed the novel SNN model which, by including an SA layer for feature embedding, significantly improves DOA estimation performance while  reducing the need for extensive training data. Rigorous feature analysis and performance evaluation confirmed the effectiveness of the proposed approach, demonstrating a substantial advancement in dynamic operation environments.

\vspace{0.5em}


\bibliographystyle{IEEEtran}

\bibliography{refs}
\end{document}